\documentclass[12pt,preprint]{aastex}

\shortauthors{Liu et al.}

\begin{document}

\title{Geometry, Kinematics and Heliospheric Impact of a Large CME-driven Shock in 2017 September} 

\author{Ying D. Liu\altaffilmark{1,2}, Bei Zhu\altaffilmark{1,2}, and Xiaowei Zhao\altaffilmark{1,2}} 

\altaffiltext{1}{State Key Laboratory of Space Weather, National Space 
Science Center, Chinese Academy of Sciences, Beijing, China; liuxying@swl.ac.cn}

\altaffiltext{2}{University of Chinese Academy of Sciences, Beijing, China}

\begin{abstract}

A powerful coronal mass ejection (CME) occurred on 2017 September 10 near the end of the declining phase of the historically weak solar cycle 24. We obtain new insights concerning the geometry and kinematics of CME-driven shocks in relation to their heliospheric impacts from the optimal, multi-spacecraft observations of the eruption. The shock, which together with the CME driver can be tracked from the early stage to the outer corona, shows a large oblate structure produced by the vast expansion of the ejecta. The expansion speeds of the shock along the radial and lateral directions are much larger than the translational speed of the shock center, all of which increase during the flare rise phase, peak slightly after the flare maximum and then decrease. The near simultaneous arrival of the CME-driven shock at the Earth and Mars, which are separated by 156.6$^{\circ}$ in longitude, is consistent with the dominance of expansion over translation observed near the Sun. The shock decayed and failed to reach STEREO A around the backward direction. Comparison between ENLIL MHD simulations and the multi-point in situ measurements indicates that the shock expansion near the Sun is crucial for determining the arrival or non-arrival and space weather impact at certain heliospheric locations. The large shock geometry and kinematics have to be taken into account and properly treated for accurate predictions of the arrival time and space weather impact of CMEs.  

\end{abstract}


\section{Introduction}

Coronal mass ejections (CMEs) are massive expulsions of plasma and magnetic flux from the solar corona \citep[][and references therein]{webb12}. Fast CMEs can drive shocks in the corona and interplanetary space, which are key accelerators of solar energetic particles \citep[SEPs;][]{reames99}. Characterizing the shock three-dimensional (3D) geometry and kinematics and their connection with heliospheric impacts is of crucial importance for space weather research and forecasting.      

CME-driven shocks have been tracked using type II radio bursts, with an effort to determine the shock distance and speed from the frequency drift of the associated type II burst \citep[e.g.,][]{reiner07, liu08, liu17b, hu16, zhao17, zucca18}. Type II emissions, in general, cannot give details of shock geometry, although radio triangulation can provide some information of where in the shock the emission originates \citep[e.g.,][]{juan12, magdalenic14}. In situ solar wind measurements have also been extensively used to study CME-driven shocks in interplanetary space \citep[e.g.,][]{burton92, mostl12, riley16, lium18}. In particular, multiple spacecraft measurements at different locations in the heliosphere have indicated a large angular extent of CME-driven shocks \citep[e.g.,][]{reisenfeld03, liu08, delucas11}. For instance, a CME-driven shock from 2001 November was observed at both the Earth and Ulysses with a latitudinal separation of 73$^{\circ}$ and longitudinal separation of 64$^{\circ}$ \citep{reisenfeld03}. Another example in 2006 December was associated with an even larger angular separation between the Earth and Ulysses, i.e., 74$^{\circ}$ in latitude and 117$^{\circ}$ in longitude \citep{liu08}. In situ measurements at different spacecraft, however, are usually sporadically distributed in the heliosphere, and opportunities with well-aligned multiple spacecraft are rare for the investigation of CME-driven shocks.   

Direct imaging of CME-driven shocks are now feasible with today's coronagraphs and heliospheric imagers. They are usually observed as a faint edge ahead of the CME front in coronagraph images near the Sun \citep[e.g.,][]{vourlidas03, vourlidas09, liu08, ontiveros09, hess14}. When the shocks are far away from the Sun, they appear as a broad front in heliospheric imaging observations \citep{liu11, liu12, liu13, maloney11, volpes15}; the density within the CME driver decreases due to expansion in interplanetary space, so the shock and sheath become dominant in heliospheric imagers. These white-light imaging observations over a large distance range allow a possible prediction of the shock parameters at 1 AU \citep[e.g.,][]{liu11, liu13, volpes15}. Simultaneous observations from different vantage points enable the determination of the shock 3D geometry and kinematics with respect to the CME driver \citep{kwon14, liu17a}. However, the 3D nature and kinematics of CME-driven shocks in relation to their heliospheric impacts, which requires coordinated remote-sensing and in situ observations from multiple spacecraft, are still not well understood.  

The 2017 September 10 eruption, which was observed by a fleet of spacecraft at different vantage points, provides a great opportunity to study the 3D geometry, kinematics and heliospheric impacts of CME-driven shocks. The event has attracted significant attention owing to its unusual energetics and occurrence near the end of a weak solar activity cycle \citep[e.g.,][and references therein]{lee18, luhmann18, seaton18, li18, hu18, yan18, gopalswamy18}. In this work we take advantage of the optimal, multi-spacecraft observations of the 2017 September 10 eruption to gain insights into CME-driven shocks. The view from the Earth as a limb event allows to track the flux rope and the CME-driven shock from their nascent stage continuously to the outer corona with projection effects minimized. In situ measurements from two well-separated locations also enable assessment of the heliospheric impacts in connection with remote-sensing observations. 

\section{Observations and Analysis} 

The eruption was associated with a long-duration X8.3 flare from NOAA AR 12673 (S09$^{\circ}$W91$^{\circ}$), which peaked at 16:06 UT on September 10. Figure~1 shows the eruption at its nascent stage viewed from GOES 16 and the well-developed CME from the Solar and Heliospheric Observatory \citep[SOHO;][]{domingo95} and the Solar Terrestrial Relations Observatory A spacecraft \citep[STEREO A;][]{kaiser08}. The combined EUV and coronagraph observations allow us to track the flux rope and shock separately from the early stage to the outer corona. The axis of the flux rope must be nearly parallel to the ecliptic plane, because otherwise the current sheet behind the flux rope would not be seen in the EUV observations. This interpretation is supported by the orientation of the post-eruption arcades (not shown here) and white-light CME reconstruction (see below). Note the wave and magnetic loop produced by the expansion of the flux rope in the EUV running-difference images. The EUV wave at the early time, which is the footprint of the CME-driven shock in the lower atmosphere \citep[e.g.,][and references therein]{patsourakos12, kwon14}, and the top of the magnetic loop, which should be very close to the shock front in the upper atmosphere, can be used together to constrain the shock structure. The shock is also seen in the coronagraph images as a faint edge around the CME. It was expanding in all directions and appeared to enclose the whole Sun in both SOHO and STEREO A images at the times given in Figure~1. Similar shock geometries with 360$^{\circ}$ envelope around the Sun have also been found in other cases \citep[e.g.,][]{kwon15, liu17a}.

Based on the observations of the 2012 July 23 complex CME, \citet[][hereinafter referred to as Paper 1]{liu17a} investigate the structure, propagation and expansion of the associated shock. They find that the shock can be modeled well by a simple spheroidal structure as in other studies \citep[e.g.,][]{hess14, kwon14}, which enables a separation between translation and expansion of the shock. Here we use an ellipsoidal model to simulate the shock \citep{kwon14, kwon15}, which can be applied to both the EUV and white-light observations. The cross section of the shock ellipsoid perpendicular to the propagation direction is assumed to be circular, which reduces a free parameter in the model. As for the CME, we employ a graduated cylindrical shell (GCS) method, which assumes a rope-like morphology with two ends anchored at the Sun, to estimate its propagation direction and distance \citep{thernisien06}. The GCS model is applied to the bright rim of the CME in coronagraph observations, for which the method is designed. Reasonable fits are generally obtained for both the CME and shock by adjusting the model parameters to visually match the two views from the Earth and STEREO A simultaneously (see Figure~1).    

Figure~2 displays the modeled CME and shock projected onto the ecliptic. The average propagation longitude of the CME and shock is about 95$^{\circ}$ and 97$^{\circ}$ west of the Earth, respectively. The CME flux-rope has a small tilt angle (about 20$^{\circ}$), consistent with what the EUV images imply. The ellipsoidal structure of the shock is produced by the vast expansion of the ejecta in combination with its outward motion, as suggested in Paper 1. In agreement with this idea, the standoff distance between the shock and driver increases from the nose towards the flank. The large extent of the shock, which encircles the whole Sun, helps explain the early detection of energetic particles at Mars and STEREO A \citep{lee18, luhmann18}. However, it is worth noting that at some point, especially near the wake, the structure in question could be just a wave without a non-linear steepening character (see Paper 1). Again, the simple structure of the shock enables an easy separation between the translational distance of the shock center and the expansion distances along and perpendicular to the propagation direction. 

The kinematics of the CME and shock are shown in Figure~3. All the speeds increase during the flare rise phase, peak slightly after the flare maximum, and then decrease. (There is a bump in the X-ray flux before the maximum, which may indicate a small flare ahead of the major one.) The maximum radial nose velocity is about 3000 km s$^{-1}$ for the CME and about 3300 km s$^{-1}$ for the shock\footnote{Note that there is a small convex-outward structure at the nose (see Figure~1), which is not covered by the present shock modeling. Its speed may be higher than given here. \citet{gopalswamy18} obtain a larger speed (about 4000 km s$^{-1}$) using a spherical shock model to cover the convex-outward structure, but other parts of the shock around the nose are not modeled well.}. The majority of the shock nose speed comes from the radial expansion speed with a peak value of about 2400 km s$^{-1}$. The radial expansion speed is always larger than the translational speed of the shock center, whose peak value is only of the order of 1000 km s$^{-1}$. The lateral expansion speed of the shock closely follows the radial expansion speed and is a little larger.

Figure~4 shows the in situ measurements at the Earth (1.01 AU) and Mars (1.66 AU). No ICME signatures were observed at the Earth, and only a forward shock arrived around 19:21 UT on September 12. A shock was also observed at Mars around 02:52 UT on September 13, possibly followed by a brief ICME interval \citep{lee18}. The shock arrival time at Mars is determined from the peak in the high-energy particles (see the bottom panel of Figure~4), known as energetic storm particles trapped around the shock. This arrival time agrees with the solar wind plasma measurements at Mars although the measurements are sparse \citep{halekas17}. The ICME and shock failed to arrive at STEREO A (not shown here). The arrival or non-arrival situation of the ICME/shock at the three locations is consistent with what the modeled CME and shock geometries suggest in Figure~2. The shock around the backward direction decayed before reaching the distance of STEREO A, like the case in Paper 1. 

Of particular interest is the near simultaneous arrival of the shock at the Earth and Mars, despite a radial separation of about 0.65 AU between them. The active region produced several eruptions, but our analysis shows that the same CME on September 10 reaching both locations is the most likely scenario. Note that the shock speed at the Earth is about 570 km s$^{-1}$, with which it would take about 2 more days to reach the distance of Mars. This near simultaneous arrival tells the importance of the shock geometry and kinematics on its impact in the heliosphere. If the shock is assumed to be spherical, the locations of the Earth and Mars relative to the shock propagation direction (see Figure~2), together with a simultaneous arrival, would give a distance of about 0.9 AU from the Sun for the shock center and a radius of about 1.43 AU for the shock sphere. This is consistent with the dominance of expansion over translation as inferred from Figure~3.   

To connect the imaging observations with the in situ measurements, we experiment with WSA-ENLIL MHD simulations \citep[][hereinafter referred to as ENLIL]{arge04, odstrcil04}. The CME is inserted in the simulation as a plasma cloud through the inner boundary at 21.5 solar radii from the Sun with CME parameters determined from a cone model \citep{zhao02}. It has no magnetic field of its own and merely carries whatever field is present from the ambient solar wind model. This limitation will to some extent determine how the ``CME" and shock evolve in the solar wind. Nevertheless it can do a reasonable job in some cases of predicting shock arrivals and jumps for example. The first run uses the measured CME half width from the cone model ($58^{\circ}$), and for the second run we increase the half width to $90^{\circ}$ to (partly) account for the large angular extent of the shock. The CME density is decreased by half for the second case, so the CME mass is roughly the same.

The simulation results are displayed in Figure~5, and the extracted time series of plasma and magnetic field parameters at the locations of the Earth and Mars are plotted in Figure~4 to compare with the actual measurements. For the first case using the measured CME width, the Earth is completely missed, and the shock arrival time at Mars is around 21:00 UT on September 13, which is about 18 hr later than observed (also see Figure~4). The simulation indicates that the Earth would only see a weak co-rotating interaction region (CIR). For the second case with the increased CME width, the shock reaches both the Earth and Mars with much better improved arrival times. The arrival time is around 16:00 UT on September 12 at the Earth, which is only about 3 hr earlier than observed, and around 10:00 UT on September 13 at Mars, which is only about 7 hr later than observed. In this case, the impact at Mars is much stronger, and the Earth sees the shock on top of the CIR. These results indicate that the shock geometry and kinematics must be properly simulated to have an accurate predication of the arrival and impact. 

Although Mars was about $60^{\circ}$ away from the CME/shock nose, the event caused significant space weather effects at Mars \citep[][and references therein]{lee18}. The large expansion of the shock observed near the Sun must have enhanced the impact at Mars, as demonstrated above. The two preceding CMEs from September 9, which are included in the simulations, might also have helped by changing the plasma and magnetic field distributions in the ejecta. The simulations indeed show asymmetric distributions in the ejecta mass and speed with respect to the CME propagation direction, part of which can be attributed to the merging of the three CMEs.

\section{Conclusions and Discussion}

We have investigated the large CME-driven shock associated with the 2017 September 10 eruption, taking advantage of the coordinated remote-sensing and in situ observations from a fleet of well-aligned spacecraft. The results, which complement the findings of Paper 1 on the structure and evolution of CME-driven shocks, shed light on the 3D geometry and kinematics of CME-driven shocks in relation to their heliospheric impacts. 

The shock and the CME driver can be tracked separately from the early stage to the outer corona using the multi-point EUV and white-light observations. The shock showed a large oblate structure and eventually enclosed the whole Sun, which can be attributed to the vast expansion of the ejecta in combination with its outward motion. The standoff distance between the shock and driver increases from the nose towards the flank and wake. The expansion speeds of the shock along the radial and lateral directions, which closely follow each other, are much larger than the translational speed of the shock center. The expansion and translational speeds increase during the flare rise phase, peak slightly after the flare maximum, and then decrease. Previous studies suggest a temporal relationship of CME acceleration with the associated flare \citep[e.g.,][]{zhang01}. Here our work indicates that a similar relationship also exists between shock kinematics and the flare. Our results also imply that the apparent acceleration of the CME leading edge, which is often used to investigate the acceleration mechanism of CMEs \citep[e.g.,][]{zhang01, bein11}, is largely dominated by the expansion due to the ejecta internal overpressure. The enormous lateral expansion of the shock has important implications for SEP production, because the shock lateral part can have a favorable geometry with respect to the ambient magnetic field for particle acceleration \citep[e.g.,][]{kozarev15, zhu18}. This geometry, together with the large lateral expansion speed, may produce enhanced SEPs even for observers connected to the flank of the shock (i.e., not the nose). This helps explain the ground level enhancement observed at both the Earth and Mars in the current case \citep{lee18, luhmann18, gopalswamy18}.

The CME-driven shock arrived at both the Earth and Mars near simultaneously, which agrees with the dominance of expansion over translation observed near the Sun. These two well-separated locations suggest a large longitudinal extent of the shock in interplanetary space (at least 156.6$^{\circ}$). This corresponds with previous multi-point in situ measurements \citep[e.g.,][]{reisenfeld03, liu08, delucas11} and here multi-spacecraft imaging observations of exceptionally wide shock structure. Note that, however, Äthe shock around the backward direction may have decayed before reaching the distance of STEREO A, since no shock signatures were observed at STEREO A. As we have found in Paper 1, at some point, especially near the wake, the structure in question could be just a wave without a non-linear steepening character; the shock would decay and disappear when the driver's influence becomes weak. An important lesson we have learned by comparing ENLIL MHD simulations with the multi-point in situ measurements is that the shock expansion near the Sun is crucial for determining the arrival or non-arrival and space weather impact at certain heliospheric locations. To improve the predictions of the arrival time and space weather impact, the large shock geometry and kinematics have to be properly treated. Purely using the measured CME width in a simulation like ENLIL may not be sufficient for space weather prediction. Note that here we have compared two cases of simulations, which have the same solar wind background; the effects of the ambient solar wind on the shock propagation and impact, if any, are already taken into account in the simulations.   

\acknowledgments The research was supported by NSFC under grant 41774179 and the Specialized Research Fund for State Key Laboratories of China. We acknowledge the use of data from SOHO, STEREO, SDO, GOES, Wind and MAVEN, thank M. Leila Mays, Christina O. Lee and Janet G. Luhmann for their help and discussions on the work, and are grateful to R.-Y. Kwon for providing his ellipsoidal shock model. The ENLIL simulations were provided by CCMC through their public Runs on Request system (\url{http://ccmc.gsfc.nasa.gov}).

\clearpage

\begin{figure}
\epsscale{0.7} \plotone{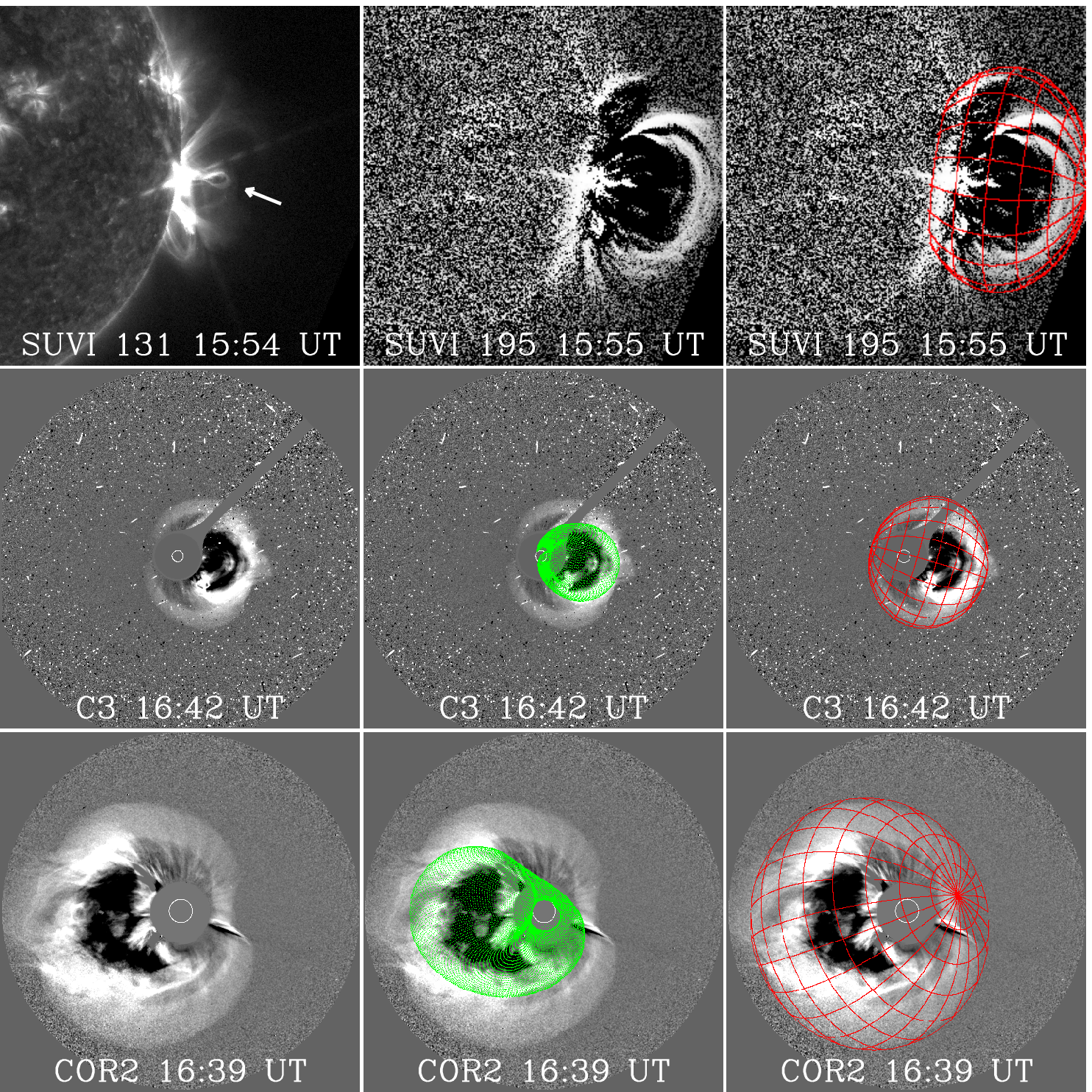} 
\caption{Multi-point imaging and modeling of the CME and shock. Top row: flux rope at 131 \AA\ from GOES 16 (left), running-difference image at 195 \AA\ (middle) and corresponding shock modeling (right). A current sheet is visible post the flux rope. Middle row: running-difference coronagraph image from SOHO (left) and corresponding modeling of the CME (middle) and shock (right). Bottom row: running-difference coronagraph image and modeling from STEREO A (128.1$^{\circ}$ east of the Earth) near the same time. Note the backward expansion of the CME-driven shock.}
\end{figure}

\clearpage

\begin{figure}
\epsscale{0.7} \plotone{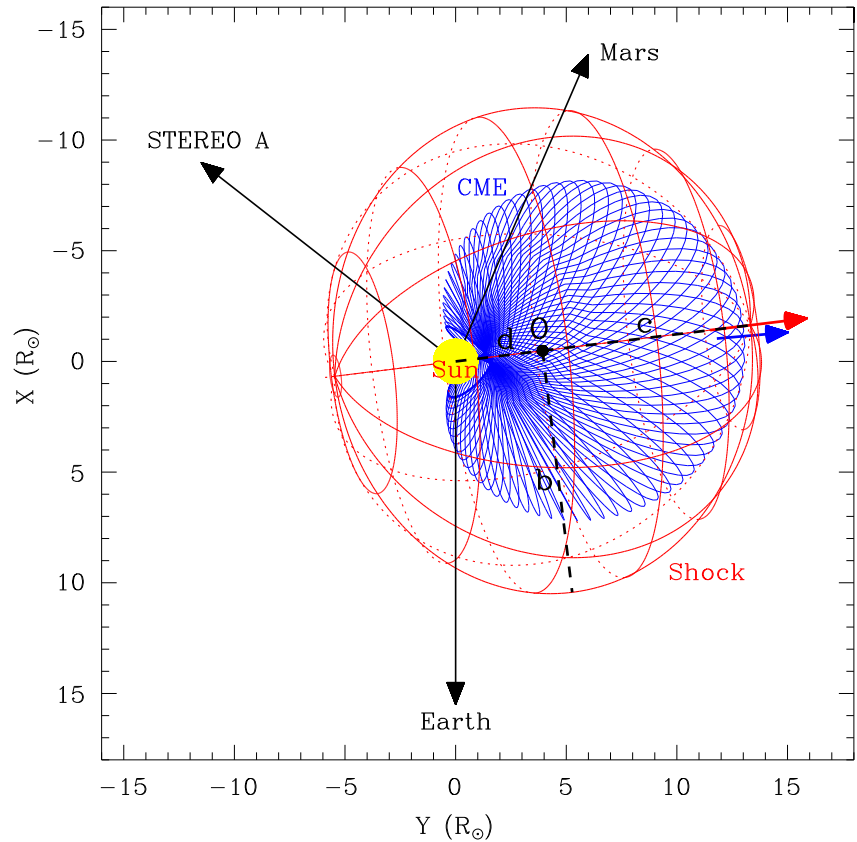} 
\caption{Projections of the modeled CME and shock onto the ecliptic plane. The directions of the Earth, STEREO A (128.1$^{\circ}$ east) and Mars (156.6$^{\circ}$ west) are shown by the black arrows. The blue and red arrows mark the propagation directions of the CME and shock, respectively. The 3D structures are taken from the modelings at 16:42 UT on September 10. Also indicated are the center of the shock ellipsoid (O), and its distances from the center of the Sun ($d$), from the nose of the shock ($c$) and from the flank perpendicular to the propagation direction ($b$).}
\end{figure}

\clearpage

\begin{figure}
\epsscale{0.65} \plotone{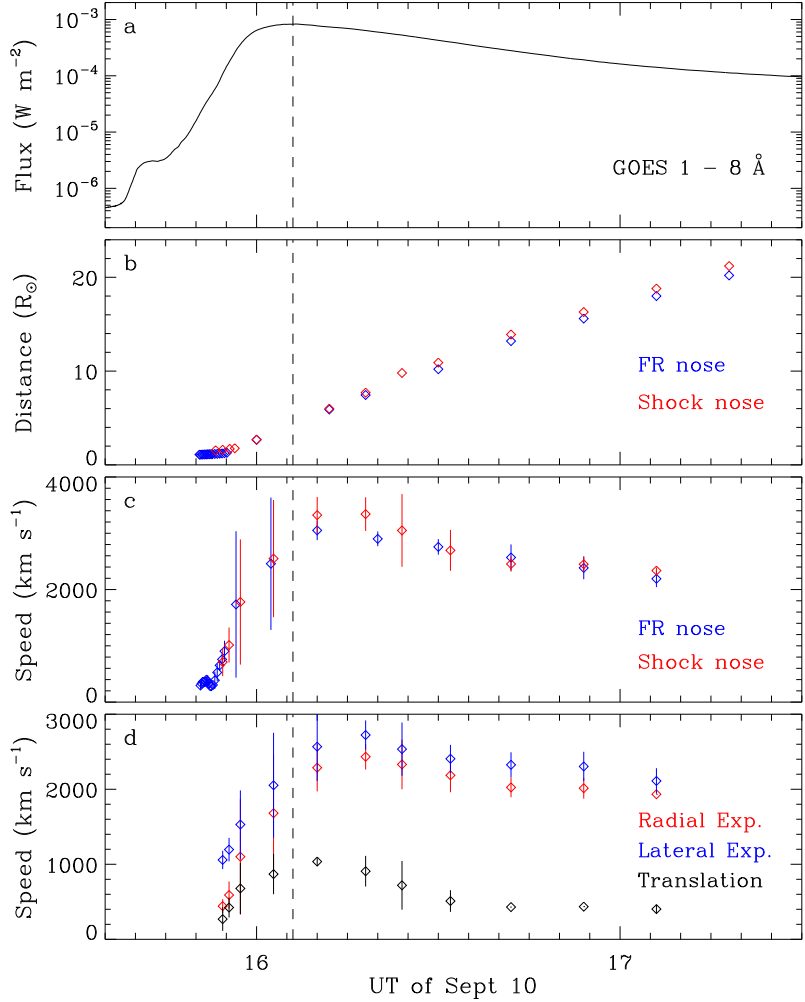} 
\caption{Kinematics of the CME and shock in comparison to the flare radiation. (a) Flare flux at 1 - 8 \AA. (b) Radial nose distances of the flux rope (FR) and shock from the center of the Sun (distances below 2 solar radii are mainly determined from EUV observations near the Earth). (c) Radial nose speeds of the CME and shock relative to the Sun. (d) Expansion speeds of the shock along the radial and lateral directions and translational speeds of the shock center. The radial expansion speed is derived from the numerical differentiation of the shock nose distance from the shock center, the lateral expansion speed is from the flank distance from the shock center, and the translational speed is from the shock center distance from the center of the Sun (see Figure~2). The vertical dashed line indicates the time of the flare maximum.}
\end{figure}

\clearpage

\begin{figure}
\epsscale{0.7} \plotone{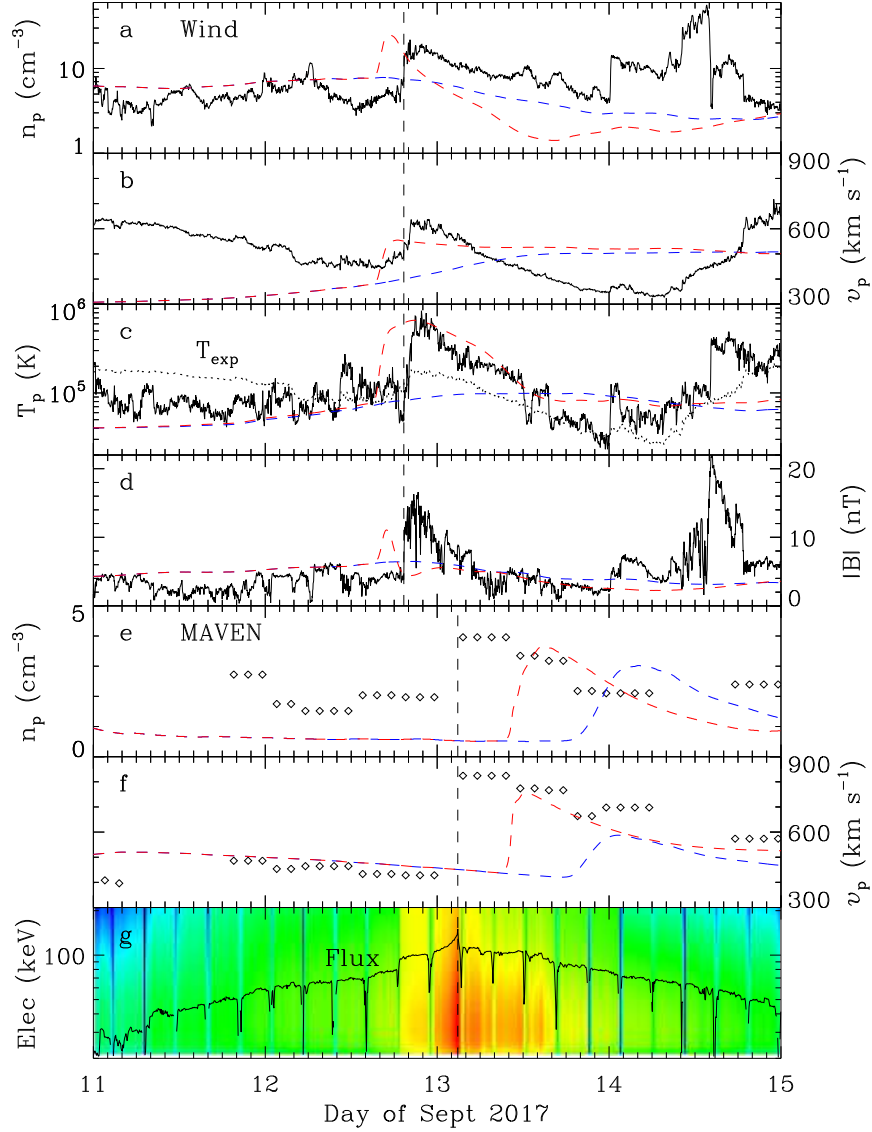} 
\caption{In situ solar wind measurements and modeling. (a-d) Proton density, bulk speed, temperature and magnetic field strength from Wind. (e-g) Proton density, bulk speed and electron differential flux (descending from red to blue) at 20-210 keV from MAVEN. The dotted curve in the third panel denotes the expected proton temperature calculated from the observed speed \citep{lopez87}. The black curve in the bottom panel represents a normalized flux averaged over the electrons. The 4.5-hour periodicity in the data comes from the orbital period of MAVEN around Mars. The vertical dashed line indicates the shock arrival at each spacecraft. The blue and red dashed curves show the simulated time series at Wind and MAVEN corresponding to the left and right cases in Figure~5, respectively.}
\end{figure}

\clearpage

\begin{figure}
\centerline{\includegraphics[width=18pc]{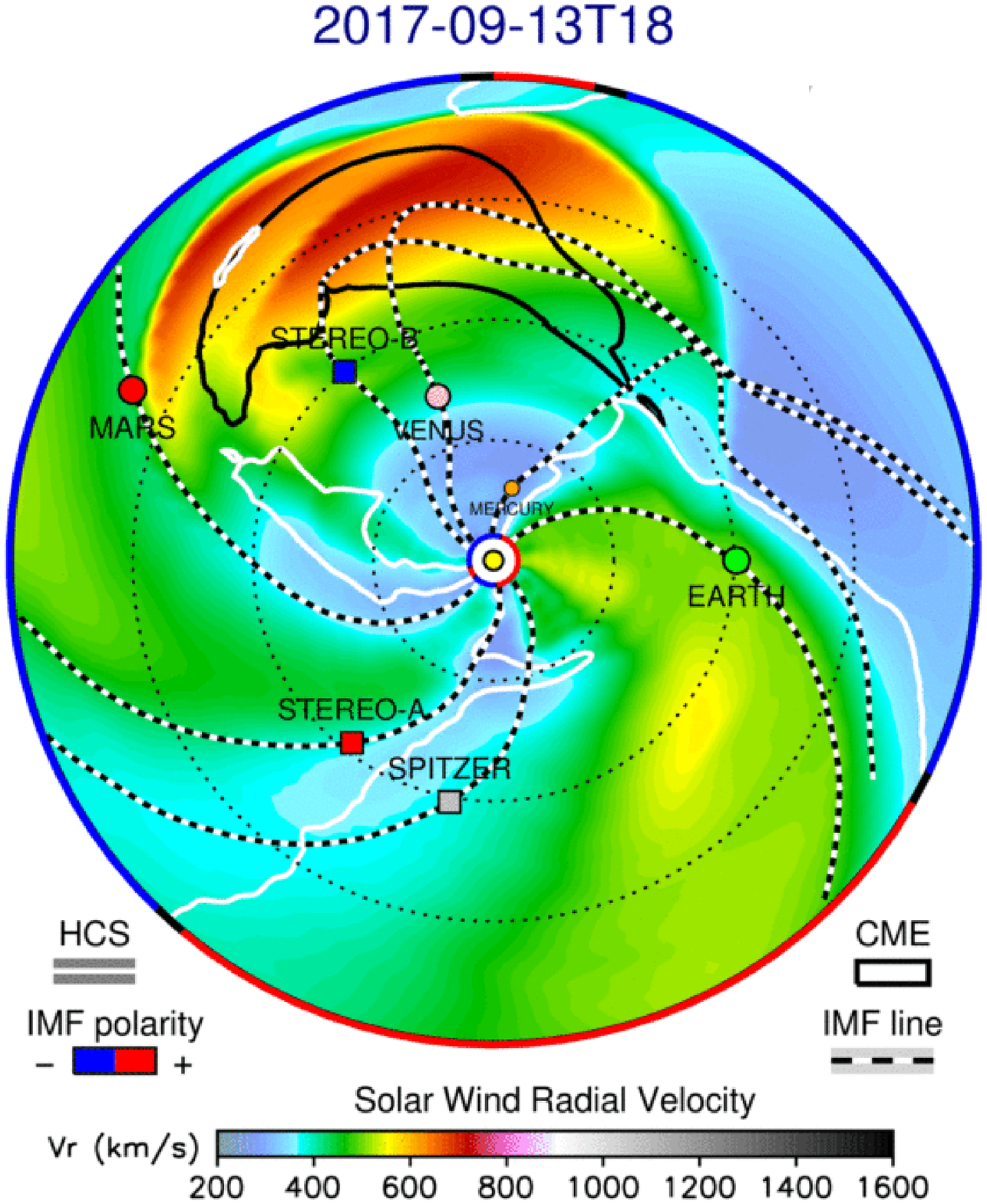}\hspace{0pc}\includegraphics[width=18pc]{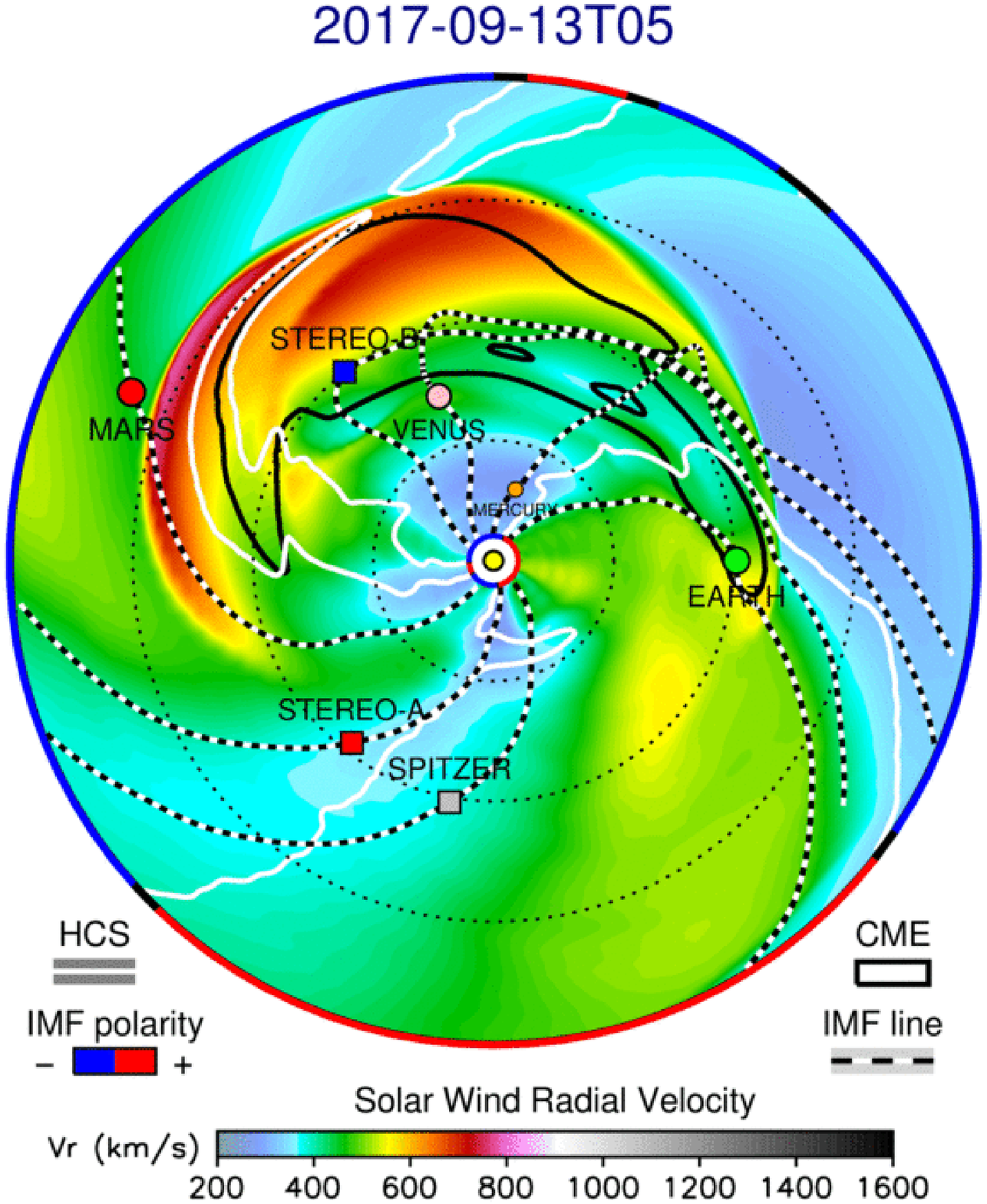}}
\caption{ENLIL MHD simulations with CME half width of $58^{\circ}$ (left) and $90^{\circ}$ (right) as input. The color shading indicates the radial solar wind speed in the ecliptic plane. Two preceding smaller CMEs from September 9 with propagation longitudes of $110^{\circ}$ and $115^{\circ}$ west of the Earth, respectively, are also included in the simulations. At the times given here, the three CMEs have already merged.}
\end{figure}

\end{document}